\documentclass[a4paper,10pt]{article}
\usepackage{url}
\usepackage{graphicx}
\usepackage{float}

\title{Reliable reconstruction of HIV-1 whole genome haplotypes reveals clonal interference and genetic hitchhiking among immune escape variants}

\author{Aridaman Pandit$^{*,1}$ and Rob J de Boer$^1$\\
$^1$Theoretical Biology and Bioinformatics, Utrecht University,\\
Utrecht, The Netherlands\\
$^*$ Email: pandit.aridaman@gmail.com}

\begin{document}
\maketitle

\begin{abstract}
Following transmission, HIV-1 evolves into a diverse population, and next generation sequencing enables us to detect variants occurring at low frequencies. Studying viral evolution at the level of whole genomes was hitherto not possible because next generation sequencing delivers relatively short reads. We here provide a proof of principle that whole HIV-1 genomes can be reliably reconstructed from short reads, and use this to study the selection of immune escape mutations at the level of whole genome haplotypes. Using realistically simulated HIV-1 populations, we demonstrate that reconstruction of complete genome haplotypes is feasible with high fidelity. We do not reconstruct all genetically distinct genomes, but each reconstructed haplotype represents one or more of the quasispecies in the HIV-1 population. We then reconstruct 30 whole genome haplotypes from published short sequence reads sampled longitudinally from a single HIV-1 infected patient. We confirm the reliability of the reconstruction by validating our predicted haplotype genes with single genome amplification sequences, and by comparing haplotype frequencies with observed epitope escape frequencies. Phylogenetic analysis shows that the HIV-1 population undergoes selection driven evolution, with successive replacement of the viral population by novel dominant strains. We demonstrate that immune escape mutants evolve in a dependent manner with various mutations hitchhiking along with others. As a consequence of this \textit{clonal interference}, selection coefficients have to be estimated for complete haplotypes and not for individual immune escapes.
\end{abstract}

\section{Background}
Inside every human host, HIV-1 embarks upon an arms race that helps it to evade the host's immune responses while maintaining its replicative fitness \cite{boutwell2010, hill2012}. Starting from a single (or a few) founder genetic lineage(s) infecting the CD4+ cells localized in the mucosa at the port of entry, HIV-1 spreads to regional lymph nodes where it comes in contact with a large number of CD4+ T cells \cite{boutwell2010, little1999, salazar2008, mcmichael2009, song2012}. This leads to rampant viral replication resulting in millions of viral particles per ml of plasma at peak viremia, occurring $\sim$21-28 days post infection \cite{little1999, fiebig2003}. Interplay between HIV-1, target cell availability, and the host's immune system during the acute phase of infection leads to a gradual decline of the viral load, eventually leveling off to a ``set point'' \cite{hill2012, mcmichael2009, goonetilleke2009}. The decrease in the viral load following peak viremia is attributed to CD4+ T cell depletion, the immune responses of HIV-1 specific B cells, and cytotoxic T cells (CTLs). A high mutation rate and a large population size allow HIV-1 to rapidly evolve immune escape mutants \cite{mansky1995, henn2012}, and strong CTL selection pressures confer fitness advantage to viral lineages harboring CTL escape mutations \cite{little1999, ganusov2006, kadolsky2010}. The efficacy and breadth of cognate CTL responses increases the within-host HIV-1 diversity after peak viremia, and plays a major role in HIV-1 evolution during the acute phase of infection \cite{mcmichael2009, goonetilleke2009, althaus2008, ganusov2011, deutekom2013, ganusov2013, liu2013}.

Longitudinally collected samples reveal the variation in the frequency of escape mutants, which can be used to estimate the selection pressure imposed by the corresponding CTL response \cite{henn2012, lee2008, da2012}. The rate of escape is typically estimated using a logistic curve, to model the replacement of one variant by another \cite{maree2000}. Several escape mutations of the same epitope may occur simultaneously in different viral lineages, but such multiple escape pathways for a single epitope are generally modeled as a single escape event \cite{ganusov2006, ganusov2013, liu2013, leviyang2013}. Escape mutations typically result in a fitness cost to the virus, and along different escape pathways the virus may face a different selection pressure by the CTLs \cite{boutwell2010, lee2008, da2012, leviyang2013, feeney2005}. Therefore, it was recommended to consider different escape pathways within the same epitope as separate escape events \cite{leviyang2013}.

Escape mutations acquired in different epitopes are typically considered independent events. But in reality multiple epitopes may escape together in either the same lineage, or in different lineages, of the viral population \cite{boutwell2010, henn2012}. Viral lineages with a high relative fitness are expected to cause a selective sweep, reducing viral diversity and driving other lineages to extinction. The concurrent presence of different epitope escapes in different viral lineages results in competition between them in their race towards fixation \cite{leviyang2013, strelkowa2012}, and this phenomenon is known as \textit{clonal interference} \cite{de1999, miralles1999}. Due to clonal interference, two beneficial mutations residing in two different lineages out-compete each other as well as the wild-type when one approaches fixation. Additional beneficial mutations tend to be acquired sequentially, ultimately resulting in the escape from multiple epitopes \cite{ganusov2013, da2012, leviyang2013}. As a result of clonal interference, epitope escape mutations may drive several non-epitope mutations, co-occurring in the same viral lineage, to fixation via \textit{genetic hitchhiking} \cite{strelkowa2012, lang2013, zanini2013}. To study the role of clonal interference and genetic hitchhiking in HIV-1 evolution, one requires long, or ideally complete, genome sequences with high coverage from multiple time points.

The availability of longitudinally sampled single genome sequences is rather limited, and next generation sequencing (NGS) is economically a more feasible alternative for single genome sequencing \cite{boutwell2010, salazar2008, fischer2010}. One large NGS dataset was generated by Henn et al. \cite{henn2012}, when they performed NGS on complete HIV-1 genomes from 6 longitudinal samples in one patient (subject 9213). NGS allowed Henn et al. \cite{henn2012} to study the dynamics of immune escape mutations in individual epitopes. However, due to the short sequencing read lengths of about $\sim$ 400bp, Henn et al. \cite{henn2012} were not able to distinguish between multiple epitopes escaping simultaneously in the same haplotype ,or independently in distinct viral haplotypes. We here attempt to reconstruct whole genome haplotypes from this data. This is challenging due to two reasons \cite{zagordi2010b, schirmer2012}. First, a variety of sequencing and PCR errors are incorporated in sequencing reads, which require robust error correction techniques to identify the true mutations \cite{zagordi2010b, topfer2013, zagordi2010}. A second problem concerns the determination of contiguity of sequencing reads generated from a single viral genome haplotype. Differences in prevalence of the genome haplotypes, genetic distances between polymorphic sites, and the amount of sequencing errors, are important factors influencing the reliability of haplotype reconstruction. A number of algorithms have been developed for viral haplotype reconstruction \cite{zagordi2010b, schirmer2012, prosperi2012, zagordi2011}, and among them PredictHaplo has been shown to perform best in reconstructing gene haplotypes at a high sequence divergence \cite{schirmer2012}.

We here demonstrate that reconstruction of whole genome HIV-1 haplotypes from NGS datasets is feasible using a read clean-up step before haplotype prediction with PredictHaplo. The reconstructed genome haplotypes provide important insights into the evolutionary dynamics of the viral quasispecies, as we show that the epitope escape dynamics are influenced by the strength and breadth of CTL selection, clonal interference, and genetic drift.

\section{Results and Discussion}
\subsection{Testing the feasibility of haplotype reconstruction using simulated datasets}
Before reconstruction of HIV-1 whole genome haplotypes from next generation sequencing reads, we first tested the feasibility of the haplotype reconstruction pipeline using \textit{in silico} datasets. We generated six \textit{in silico} HIV-1 populations with realistic nucleotide diversities (2, 4 and 10\%) \cite{henn2012, burger1991, keele2008} and frequency distributions (uniform or log-normal, see Methods). For each dataset, ART\_454 software \cite{Huang2012} was used to generate \textit{in silico} next-generation sequencing reads. Each simulated population consisted of 9 master HIV-1 genomes that were 8800 bp long. For the 3 data sets with uniform frequency distributions (data sets U2, U4 and U10, respectively), our pipeline reconstructed 9 haplotypes from the short reads artificially generated from the true 9 \textit{in silico} genomes (see Methods). The reconstruction was accurate with a Hamming distance of just 1 to 9 nucleotides from the true genomes (Table 1). The accuracy decreased when the population structure was changed to a log-normal frequency distribution (data sets L2, L4 and L10, respectively). For instance, for the L4 dataset with a 4\% nucleotide diversity, the 9 reconstructed haplotypes had a Hamming distance of 1 to 427 nucleotides from the corresponding true genomes (Table 1). Closer examination, however, revealed that the four most prevalent ($>$4\%) haplotypes were accurately reconstructed, with Hamming distances of 1, 1, 19 and 23. Additionally, most of the errors were limited to either the start or end of the genomes, where the simulated coverage was low. Ignoring the initial and final 20 bases of the haplotype sequences, the four most prevalent haplotypes were identical to their closest true genomes. Finally, the genotype frequencies of predicted haplotypes matched those of the true genomes (results not shown). Summarizing, genomes prevalent at a $>$4\% frequency can be reconstructed with high fidelity, whereas reconstruction of low frequency ($<$4\%) genome haplotypes is unreliable.

In reality, HIV-1 is known to have a complex population structure involving multiple co-occurring quasispecies \cite{holmes2002}. To mimic this, we simulated a more complex population (\textit{LQ4\_i}) consisting of 9 master HIV-1 genomes corresponding to 9 quasispecies, each surrounded by a varying number of closely related genomes (Table 2 and Figure 1A; see Methods). This was simulated 10 times, and for each simulation at least 6 haplotypes were reconstructed (Table 2). The 4 most abundant haplotypes represented the 4 most prevalent quasispecies in the population, and these were more accurately reconstructed than the low frequency haplotypes. Most sequence errors were found in the initial and final 50 bp of the reconstructed haplotypes, exclusion of which resulted in a large reduction in the Hamming distance between the reconstructed haplotypes and the true genomes, especially for the prevalent haplotypes (Table 2).

We now present a detailed analysis of LQ4\_1, as a representative example (the 9 other simulations were similar). The five most prevalent haplotypes (haplo\_0, haplo\_3, haplo\_4, haplo\_5 and haplo\_6; see Figure 1) were reconstructed at a Hamming distance of 8 to 46 nucleotides from their closest sequences (Table 2), and were reconstructed as a ``consensus genome'' of their corresponding quasispecies (Figure S1). However, the most abundant quasispecies containing 50 simulated genomes, was reconstructed as two haplotypes, i.e. haplo\_0 and haplo\_4, each ``representing'' 22 and 28 simulated genomes (Figure 1A). The remaining two haplotypes (haplo\_1 and haplo\_2) were reconstructed as consensus genomes of the five rare (frequencies $<$4\%) quasispecies (Figure 1A). Variation in the Hamming distance between the reconstructed haplotype and genomes of the corresponding quasispecies was comparable to the intrinsic variation within the quasispecies (Figure 1B). The dominant quasispecies were accurately reconstructed, but the less frequent quasispecies were reconstructed at a much higher Hamming distance (Figure 1B). Since haplotypes are reconstructed as a consensus of each quasispecies (Figure S1), variations between the reconstructed haplotypes and their true genome can at least partly be attributed to the viral quasispecies population structure. The predicted prevalence of the dominant haplotypes was comparable to the true frequencies of the corresponding quasispecies (Figure 1C). Summarizing, these simulations suggest that the haplotype reconstruction pipeline works with high fidelity for complete HIV-1 genome haplotype reconstruction using 454 sequencing reads with realistic nucleotide diversities of 2 to 10\%, and genotype frequencies larger than 4\%.

\subsection{Experimental validation of the reconstructed haplotypes}
The same methodology is now applied to the NGS data from a patient described in Henn et al. \cite{henn2012}. In brief, this patient presented during acute infection with a viral load of $9.3 \times 10^{6}$ copies/ml, and henceforth this is referred to as day 0. Six longitudinal serum samples were collected at day 0, 3, 59, 165, 476 and 1543 post presentation by Henn et al. \cite{henn2012}. The peak viral load was observed at day 3. The first four samples represent the acute phase dynamics, while the last two samples fall in the chronic phase of infection. Deep sequencing (Roche 454 Genome Sequencer FLX Titanium) was performed on each sample using four overlapping PCR amplicons spanning the complete protein coding HIV-1 genome. The average fold sequence coverage for the samples from day 0, 3, 59, 165, 476 and 1543 was 667.7, 724.4, 750.5, 299.7, 227.6 and 540.7, respectively. CTL epitope escapes restricted by subject's HLA alleles (A01, A24, B38, B44 and Cw04) were studied using a local read analysis, and confirmed using IFN-gamma ELISPOT assays. The most dominant CTL responses at day 59 were directed against the Nef-RW8 and Vif-WI9 epitopes. To validate the sequencing reads, Single Genome Amplification (SGA) was performed for \textit{vif} gene sequences from day 59. The public availability of this data allowed us to test the validity of our whole genome haplotype reconstruction pipeline.

\subsubsection{Validation of reconstructed haplotypes by SGA}
As the first validation step, the 95 SGA \textit{vif} sequences from day 59 \cite{henn2012} were aligned with the 6 \textit{vif} gene sequences extracted from the whole genome haplotypes reconstructed using the NGS data from day 59. Phylogenetic analysis showed that five reconstructed \textit{vif} sequences were identical to 48 out of 95 SGA \textit{vif} sequences (Figure 2A). Of the remaining SGA sequences, 11 were found in clades that were represented by at least one reconstructed haplotype (e.g., B7, B42, B57, B81 and B130 were represented by d59\_1). 16 low frequency SGA variants, depicted as singletons in Figure 2A, were represented by one haplotype (d59\_2). Thus, 75 of the 95 observed SGA sequences were represented by 6 reconstructed genome haplotypes. The remaining 20 SGA sequences were prevalent at low frequencies, and were not reconstructed as a unique haplotype. Nevertheless, this phylogenetic analysis shows that most of the variation in \textit{vif} is accurately represented, when reconstructing whole genome haplotypes (Figure 2A).

\subsubsection{Temporal variations in haplotype prevalence frequencies}
Next to reconstructing the sequences, we should also be able to predict the correct genotype frequencies of the haplotypes. Due to transmission bottlenecks, HIV-1 infection is typically established by a single (or a few) genetic lineage(s) \cite{boutwell2010, salazar2008, mcmichael2009, song2012, leviyang2013}, and the HIV-1 population does not exhibit much diversity until the peak viremia \cite{henn2012}. One should therefore expect one (or a few) dominant HIV-1 haplotype(s) before and around the time viremia peaks. Around the peak viremia, immune selection leads to diversification, i.e. a decrease in the frequency of the dominant quasispecies, and an increase in frequency of other quasispecies in the population \cite{boutwell2010, leviyang2013}. Reconstructing the haplotypes from every sample, we found that the number of reconstructed haplotypes varied from 3 to 8 (Figure 2B). Although the viral population at day 0 (i.e., before peak viremia) consisted of 4 haplotypes, only one haplotype (d0\_1) was present at a high frequency ($\sim$87\%), confirming that this infection was established by a single founder virus lineage. At day 3 (around peak viremia), the same dominant haplotype (d3\_0, identical to d0\_1) was present at a similar high frequency ($\sim$83.6\%). Diversification led to a decrease in the frequency of dominant haplotypes at day 59 and 165. Several haplotypes with comparable frequencies were reconstructed for day 476 and 1543, indicating diversification and expansion of multiple lineages during the early chronic phase of infection (Figure 2B). The reconstructed haplotypes thus suggest a realistic scenario where the infection in subject 9213 was started by a single haplotype, followed by an increase in viral diversity resulting in the establishment of multiple co-dominant viral quasispecies.

\subsubsection{Validation of whole genome haplotypes using predicted epitope frequencies}
As a third validation, we compared the frequencies of the CTL epitopes, obtained by summation of the predicted frequencies of those haplotypes containing the CTL epitope, with the observed read frequencies described in Henn et al. \cite{henn2012}. The average nucleotide diversity between whole genome haplotypes from the same sample was 128 nucleotides. As the epitope escape mutations comprise only a small fraction of all the polymorphic sites used to infer the whole genome haplotypes, this comparison is a fairly independent validation. Since it is trivial that the epitopes containing no variation in a given sample will always exhibit 100\% read and haplotype frequencies, they were not considered for the correlation analysis. The predicted frequencies of the epitopes were accurate and tightly correlated with the observed frequencies (Spearman's $\rho=0.85$, p-value $< 10^{-15}$) (Figure 2C). Some of the low frequency epitope variants were not predicted by haplotype reconstruction, and they could either be sequencing errors, or true variants not containing sufficient contiguous reads to allow their reconstruction. The Nef RW8-T5M escape mutation, which was not detected by the local read analysis of Henn et al. \cite{henn2012}, was predicted at a high frequency ($\sim$64\%) in the day 59 haplotypes (red circle on vertical axis, Figure 2C). Henn et al. \cite{henn2012} performed several additional clean-up steps (read phasing, read profiling, and removal of reads partially spanning an epitope), which apparently discarded the reads containing this Nef RW8-T5M variant. Because the Nef RW8-T5M mutation was present at subsequent time points \cite{henn2012}, we think its detection by the pipeline is correct. To summarize, we can correctly predict the epitope frequencies by reconstructing whole genome haplotypes.

\subsection{Biological Results}
\subsubsection{Phylogenetic analysis of reconstructed haplotypes}
A hallmark of HIV-1 evolution during the acute phase of infection is the selection by the immune system \cite{boutwell2010, grenfell2004}. The successive replacement of viral haplotypes in the phylogenetic trees in Figure 3 reveals that HIV-1 in subject 9213 also underwent a selection driven evolution. The \textit{nef}, and \textit{vif} genes, which are targeted by dominant CTL responses \cite{henn2012}, and the \textit{env} gene targeted by 3 sub-dominant CTL responses \cite{henn2012}, exhibit a strong temporal phylogenetic signal of successive replacement of dominant haplotypes (Figure 3A-C). Strong temporal selection causes all day 1543 haplotypes to form a monophyletic clade diverging from the clade containing the dominant haplotype at day 476 (d476\_1) (Figure 3A-C). The \textit{gag}, and \textit{pol} genes targeted by sub-dominant CTL responses exhibit intermediate phylogenetic signals for temporal selection (Figure 3D and E). The trees of the remaining genes (\textit{rev}, \textit{tat}, \textit{vpr}, and \textit{vpu}) did not reveal immune selection driven evolution (Figure 3F-I). The complete genome phylogeny captured the temporal dynamics exhibited by the genes under identified CTL selection, and showed similar successive replacement of dominant haplotypes (Figure 4). Thus, phylogenetic analysis indicates temporal replacement of the dominant quasispecies, and that this signal is most evident in the genes targeted by the CTLs.

\subsubsection{Clonal interference and epitope escape}
Studying the 7 identified CTL epitopes in the whole genome haplotypes reveals that the virus explores multiple escape pathways in most epitopes. Different combinations of escape mutations were found in different haplotypes from the same population (Figure 5). The presence of different combinations of escape mutations in each haplotype should confer different fitness advantages to them resulting in clonal interference between the viral haplotypes. This illustrates that the selective advantage of escape mutations has to be considered in combination. The population at day 59 contained 6 escape mutations distributed over 4 different epitopes, 2 in Nef A24-RW8, 2 in Vif B38-WI9, 1 in Gag A01-GY9 and 1 in Env A01-RY9 (Figure 5). Interestingly, haplotype d59\_3 containing 2 of the CTL escape mutations disappeared before day 165, while d59\_5, which was the dominant haplotype at day 59 with only one escape mutation survived, and additionally evolved an anchor position escape mutation in the Vif B38-WI9 epitope by day 165. Interestingly, haplotype d165\_3 contained 4 epitope escapes and was not selected over time, whereas haplotype d165\_0 containing two escape mutations was selected (Figure 5).

An epitope typically contains two anchor positions that are essential for binding the MHC molecule. An epitope with an anchor position escape therefore results in the evasion from all potential CTL responses, as the epitope is no longer presented by the MHC molecule. This should confer a higher fitness compared to non-anchor position escapes which can be targeted by other CTL populations. This could explain why haplotype d165\_0 with an anchor position escape (I9V) for the dominant Vif epitope was selected over haplotype d165\_3 with the non-anchor S8A escape (Figure 5). Additionally, the strength of CTL selection may influence the fate of other epitope escape mutations. The fixation of haplotype d165\_0, containing the escape mutations Vif B38-I9V and Nef A24-T5M from two dominant CTL responses, could therefore explain why the Gag A01-R6K and Env A01-I6V escape mutations in the d165\_2 haplotype disappeared from the population (Figure 5). Thus, several factors like the strength of each CTL response, breadth of CTL responses, and genetic drift, together appear to determine the relative fitness of a haplotype. Most importantly, beneficial epitope escape mutations disappear due to higher relative fitness of other haplotypes present in the viral population, emphasizing that clonal interference plays an important role in acquisition of epitope escape mutations by HIV-1.

\subsubsection{Selection of haplotype genomes}
Figure 6 depicts the temporal dynamics of the reconstructed HIV-1 haplotype genomes. The dominant (black circle) and other less abundant (gray squares) haplotypes at each time point were connected to the preceding haplotype with the minimal non-synonymous Hamming distance (mentioned above each line in Figure 6). In all six samples, the dominant quasispecies seeded the subsequent dominant quasispecies that was generally located at a lower mutational distance, whereas the other less prevalent quasispecies were located at a higher mutational distance from the ancestral haplotype (Figure 6). The rate at which the dominant haplotype is replaced by the subsequent haplotype (denoted by red dashed lines in Figure 6) decreased over time from 0.17 to 0.005 per day (Figure 6, inset). The rates of replacement calculated using whole genome haplotypes better represent the evolution of HIV-1 under immune selection than the set of escape rates estimated on the basis of individual epitopes \cite{henn2012}.

\subsubsection{Genetic hitchhiking}
As a consequence of clonal interference, the fixation of genomes with high relative fitness due to the presence of beneficial epitope escape mutations, may result in genetic hitchhiking of other mutations present in the same haplotype \cite{strelkowa2012, lang2013, zanini2013}. The linkage disequilibrium between the nucleotide positions in HIV-1 are shown as a network of sites that are genetically linked in the set of all 30 haplotype genome sequences (Figure 7). Different colors represent different HIV-1 genes and triangles denote the mutations in epitopes (Figure 7). There are two major clusters of linked sites, one containing sites from the \textit{gag} gene only (cluster I), and another containing sites from several genes (cluster II). In addition there are several clusters containing less than 5 sites. Epitope escapes are typically linked with one or more sites in non-epitope regions and these links could reflect compensatory mutations (Figure 7). Interestingly, the two early epitope escapes from the dominant CTL responses against Nef A24-RW8 (T5M genomic site: 8495, green arrow) and Vif B38-WI9 (genomic site: 4544 and 4547, blue and red arrow) were linked only with very few other sites (Figure 7). In contrast, the late epitope escapes from sub-dominant CTL responses like \textit{pol} and \textit{env} were genetically linked with several sites (cluster II, Figure 7).

The difference in linkage profiles can be explained by the homogeneity of the HIV-1 population at the start of infection (Figure 2B). The effect of genetic hitchhiking is clearly demonstrated by the large linkage cluster comprised of links between sites from \textit{gag} (red) to \textit{nef} (green) which probably have little or no functional linkage (cluster II, Figure 7). The hitchhiking mutations can be mildly advantageous or neutral \cite{lang2013}, or may even be potentially deleterious \cite{zanini2013}. For example, 22 non-synonymous mutations and 2 synonymous mutations (probably neutral) in the \textit{env} gene have hitchhiked with 1 beneficial \textit{env} epitope escape mutation (purple circles in cluster II, Figure 7). Thus, several non-epitope mutations hitchhike with immune escape mutations.

The haplotype reconstruction pipeline, using a combination of read clean-up and PredictHaplo, helped us to study the epitope escape at a whole genome perspective, highlighting the importance of clonal interference and genetic hitchhiking. A major drawback of this methodology may be that we miss low frequency recombinants in the population. Recombination during HIV-1 replication may facilitate the accumulation of multiple epitope escapes in single genomes. But during acute infection, the rate of recombination is known to be rather low \cite{kessinger2013, neher2010, batorsky2011}, suggesting that most epitope escape mutations are sequentially acquired by the quasispecies \cite{kessinger2013}. Note that recombinant haplotypes that become sufficiently prevalent in the population will be detected as novel unique haplotypes.

\section{Conclusions}
We have shown that whole genome HIV-1 haplotypes can be reconstructed from short 454-sequencing reads with high fidelity, albeit with errors in the first and last 50 bases, and that the predicted population dynamics, and epitope frequencies from the whole genome haplotypes are realistic. Reconstruction of genome haplotypes provides an opportunity to study the interaction between epitope escapes. Studying the complete genome haplotypes from one HLA-typed patient (subject 9213), we show that after the establishment of infection by a single viral haplotype HIV-1 diversifies considerably, leading to the co-existence of several viral lineages. Phylogenetic analysis shows that the HIV-1 population exhibits selection driven evolution, where dominant quasispecies seed the subsequent viral population. Several epitope escapes evolve in single haplotypes, and different combinations of multiple epitope escapes become prevalent in the viral population. As a result of clonal interference, the fate of each escape mutation, also depends upon the fitness of other immune escape mutations prevalent in the viral population. The long range linkage disequilibrium between genomic sites suggests that clonal interference between HIV-1 genomes results in the fixation of several non-epitope mutations via genetic hitchhiking.

\section{Methods}
\subsection{Simulated datasets generated for validation}
To validate the haplotype reconstruction pipeline, we created multiple simulated HIV-1 populations. For each of these simulated datasets, we first generated 9 mutated haplotype genomes from a reference HIV-1 genome (GenBank accession number: JQ403055) by randomly selecting $r$ sites such that $n/2\le r\le n$, where $n$ represents either 2, 4 or 10\% sites of the total number of sites. Three data sets U2, U4 and U10 (with varying nucleotide diversities of 2, 4 or 10\%) were made with a uniform frequency distribution, i.e. all haplotypes were present at the same frequency (1/9) in the population. Three other data sets were made with a log-normal frequency distribution, with the same nucleotide diversities: 2, 4 and 10\% for L2, L4 and L10, respectively. To implement a log-normal frequency distribution, some haplotype genomes were duplicated multiple times in the population to increase their frequencies. Thus, a typical population with a log-normal distribution consisted of 9 haplotypes present at the following copy numbers: 20, 12, 4, 2, 1, 1, 1, 1, 1, respectively. This generated a quasispecies with 5 minority haplotypes present at a frequency below $2.5\%$, and one dominant haplotype with a prevalence exceeding $> 45\%$.

Additionally, we simulated 10 populations with a quasispecies structure. For each ``quasispecies data set'' (LQ4\_i), we generated nine master genomes at a 4\% diversity (similar to those above). After creating the 9 master genomes, we created mutated copies with at most $0.1\%$ nucleotide diversity from the corresponding master sequence. This created a cloud of genomes around each master genome. The frequency of the nine quasispecies were distributed log-normally in the LQ4 dataset at the following copy numbers: 50, 24, 8, 4, 2, 2, 2, 2, 1 (as shown in Figure 1A).

We used the ART\_454 software \cite{Huang2012} to simulate 454 sequencing errors and generate \textit{in silico} reads with an average coverage of 500 reads per base, mimicking the average depth obtained in Henn et al. \cite{henn2012}. The \textit{in silico} reads obtained were analyzed using the haplotype reconstruction pipeline. A lower average coverage of 200 reads hardly changed the number of reconstructed haplotypes (results not shown).

\subsection{Longitudinal dataset used to study the within-host dynamics of HIV-1 haplotypes}
The 454 sequence reads from six datasets V4137, V4136, V4139, V4140, V4676 and V4678 sampled from subject 9213 at day 0, 3, 59, 165, 476 and 1543 were downloaded from the NCBI SRA database \url{http://www.ncbi.nlm.nih.gov/Traces/sra}. As described in Henn et al. \cite{henn2012}, the sequences were obtained by amplifying four overlapping PCR amplicons which span the entire protein coding region of HIV-1 genome. Subject 9213 was HLA-typed, and expressed the A01, A24, B38, B44 and Cw04 alleles. Henn et al. \cite{henn2012} recognized dominant and sub-dominant CTL responses against 7 HIV-1 epitopes in subject 9213 and described the immune escapes. We study the dynamics of these 7 escape mutations using the reconstructed haplotype genomes.

\subsection{Haplotype reconstruction pipeline}
The haplotype reconstruction pipeline consists of two major steps: 1) Read clean-up, and 2) Prediction of haplotypes. Reads obtained from 454 NGS are known to contain three major types of process errors: a) carry forward and incomplete extension errors, b) homopolymer miscall errors, and c) InDels in non-homopolymer regions \cite{henn2012, zagordi2010}. These 454 sequencer specific errors were corrected using the default parameter settings of ReadClean454 v1.0 (or RC454) software which uses \textit{Mosaik} aligner \url{http://code.google.com/p/mosaik-aligner/} to align sequences to a reference genome sequence \cite{henn2012}. The RC454 cleaned reads were then subjected to PredictHaplo (version 0.5) \url{http://bmda.cs.unibas.ch/{HIV}HaploTyper/index.html} for haplotype reconstruction. We used the sample's consensus assembly sequence obtained from Henn et al. \cite{henn2012}, as the reference genome sequence for RC454 and PredictHaplo. We tested haplotype reconstruction at different parameter settings for PredictHaplo, and the default settings gave the most reliable results (results not shown). The cleaned-up reads obtained with the RC454 package were made compatible with PredictHaplo using customized Python scripts.

\subsection{Phylogenetic Analysis}
For phylogenetic analysis, genes and genomes from the reconstructed haplotypes were aligned using ClustalW \cite{thompson2002}. The complete genome, \textit{pol}, \textit{env} and \textit{gag} alignments, used in Figure 3, were 8878bp, 3040bp, 2645bp and 1546bp long, respectively. Phylogenetic trees were reconstructed using the maximum likelihood (ML) method performed by PHYML (version 3.0) \cite{guindon2003}. ML analysis was conducted using the general time reversible (GTR) model, with six substitution rate categories for the gene phylogenies, and nine for the complete genome phylogeny, with the proportion of invariant sites set to 0. The gamma distribution parameter and nucleotide frequencies were estimated from the dataset for the phylogenetic reconstruction. From the parsimonious starting tree, the best tree generated using both the nearest neighbor interchange (NNI) and the subtree pruning and regrafting (SPR) approaches is shown in Figure 3. 100 bootstrap replicates were performed to assess the support for the branches and bootstrap support values $\ge90$ are indicated on the branches. For Figure 1A and Figure 2A, neighbor-joining phylogenetic trees were reconstructed using ClustalW because of their high sequence similarity \cite{thompson2002}.

\subsection{Selection and Linkage analysis}
To estimate the fitness and selection coefficients of mutated viral genomes, we used the method described by Maree et al. \cite{maree2000}. To estimate selection coefficients, we considered that the frequency of a mutant haplotype rises from a value of 0.03 (frequency threshold below which we did not detect any haplotype) at $t-1$ to the observed frequencies at time $t$ replacing its predecessor \url{http://bioinformatics.bio.uu.nl/rdb/fitness.html}.

Linkage disequilibrium between all polymorphic genomic sites was calculated using DNAsp and linkage values only significant after Bonferroni correction (p-value $< 0.4 \times 10^{-7}$) were considered \cite{librado2009}. Linked sites were plotted as a network using Cytoscape (version 2.8) \cite{smoot2011}.

\bibliography{Pandit_deBoer}

\begin{thebibliography}{10}

\bibitem{boutwell2010}
Christian~L Boutwell, Morgane~M Rolland, Joshua~T Herbeck, James~I Mullins, and
  Todd~M Allen.
\newblock Viral evolution and escape during acute {HIV}-1 infection.
\newblock {\em The Journal of infectious diseases}, 202(Suppl 2):S309, 2010.

\bibitem{hill2012}
Alison~L Hill, Daniel~IS Rosenbloom, and Martin~A Nowak.
\newblock Evolutionary dynamics of {HIV} at multiple spatial and temporal
  scales.
\newblock {\em Journal of molecular medicine}, 90(5):543--561, 2012.

\bibitem{little1999}
Susan~J Little, Angela~R McLean, Celsa~A Spina, Douglas~D Richman, and Diane~V
  Havlir.
\newblock Viral dynamics of acute {HIV}-1 infection.
\newblock {\em The Journal of experimental medicine}, 190(6):841--850, 1999.

\bibitem{salazar2008}
Jesus~F Salazar-Gonzalez, Elizabeth Bailes, Kimmy~T Pham, Maria~G Salazar,
  M~Brad Guffey, Brandon~F Keele, Cynthia~A Derdeyn, Paul Farmer, Eric Hunter,
  Susan Allen, et~al.
\newblock Deciphering human immunodeficiency virus type 1 transmission and
  early envelope diversification by single-genome amplification and sequencing.
\newblock {\em Journal of virology}, 82(8):3952--3970, 2008.

\bibitem{mcmichael2009}
Andrew~J McMichael, Persephone Borrow, Georgia~D Tomaras, Nilu Goonetilleke,
  and Barton~F Haynes.
\newblock The immune response during acute {HIV}-1 infection: clues for vaccine
  development.
\newblock {\em Nature Reviews Immunology}, 10(1):11--23, 2009.

\bibitem{song2012}
Hongshuo Song, Jeffrey~W Pavlicek, Fangping Cai, Tanmoy Bhattacharya, Hui Li,
  Shilpa~S Iyer, Katharine~J Bar, Julie~M Decker, Nilu Goonetilleke, Michael~KP
  Liu, et~al.
\newblock Impact of immune escape mutations on {HIV}-1 fitness in the context
  of the cognate transmitted/founder genome.
\newblock {\em Retrovirology}, 9(1):1--14, 2012.

\bibitem{fiebig2003}
Eberhard~W Fiebig, David~J Wright, Bhupat~D Rawal, Patricia~E Garrett,
  Richard~T Schumacher, Lorraine Peddada, Charles Heldebrant, Richard Smith,
  Andrew Conrad, Steven~H Kleinman, et~al.
\newblock Dynamics of {HIV} viremia and antibody seroconversion in plasma
  donors: implications for diagnosis and staging of primary {HIV} infection.
\newblock {\em Aids}, 17(13):1871--1879, 2003.

\bibitem{goonetilleke2009}
Nilu Goonetilleke, Michael~KP Liu, Jesus~F Salazar-Gonzalez, Guido Ferrari,
  Elena Giorgi, Vitaly~V Ganusov, Brandon~F Keele, Gerald~H Learn, Emma~L
  Turnbull, Maria~G Salazar, et~al.
\newblock The first {T} cell response to transmitted/founder virus contributes
  to the control of acute viremia in {HIV}-1 infection.
\newblock {\em The Journal of experimental medicine}, 206(6):1253--1272, 2009.

\bibitem{mansky1995}
Louis~M Mansky and Howard~M Temin.
\newblock Lower in vivo mutation rate of human immunodeficiency virus type 1
  than that predicted from the fidelity of purified reverse transcriptase.
\newblock {\em Journal of virology}, 69(8):5087--5094, 1995.

\bibitem{henn2012}
Matthew~R Henn, Christian~L Boutwell, Patrick Charlebois, Niall~J Lennon,
  Karen~A Power, Alexander~R Macalalad, Aaron~M Berlin, Christine~M Malboeuf,
  Elizabeth~M Ryan, Sante Gnerre, et~al.
\newblock Whole genome deep sequencing of {HIV}-1 reveals the impact of early
  minor variants upon immune recognition during acute infection.
\newblock {\em PLoS pathogens}, 8(3):e1002529, 2012.

\bibitem{ganusov2006}
Vitaly~V Ganusov and Rob~J De~Boer.
\newblock Estimating costs and benefits of {CTL} escape mutations in
  {SIV}/{HIV} infection.
\newblock {\em PLoS computational biology}, 2(3):e24, 2006.

\bibitem{kadolsky2010}
Ulrich~D Kadolsky and Becca Asquith.
\newblock Quantifying the impact of human immunodeficiency virus-1 escape from
  cytotoxic t-lymphocytes.
\newblock {\em PLoS computational biology}, 6(11):e1000981, 2010.

\bibitem{althaus2008}
Christian~L Althaus and Rob~J De~Boer.
\newblock Dynamics of immune escape during {HIV}/{SIV} infection.
\newblock {\em PLoS computational biology}, 4(7):e1000103, 2008.

\bibitem{ganusov2011}
Vitaly~V Ganusov, Nilu Goonetilleke, Michael~KP Liu, Guido Ferrari, George~M
  Shaw, Andrew~J McMichael, Persephone Borrow, Bette~T Korber, and Alan~S
  Perelson.
\newblock Fitness costs and diversity of the cytotoxic {T} lymphocyte ({CTL})
  response determine the rate of {CTL} escape during acute and chronic phases
  of {HIV} infection.
\newblock {\em Journal of virology}, 85(20):10518--10528, 2011.

\bibitem{deutekom2013}
Hanneke W.~M. van Deutekom, Gilles Wijnker, and Rob~J. de~Boer.
\newblock The rate of immune escape vanishes when multiple immune responses
  control a {HIV} infection.
\newblock {\em The Journal of Immunology}, 2013.

\bibitem{ganusov2013}
Vitaly~V Ganusov, Richard~A Neher, and Alan~S Perelson.
\newblock Mathematical modeling of escape of {HIV} from cytotoxic {T}
  lymphocyte responses.
\newblock {\em Journal of Statistical Mechanics: Theory and Experiment},
  2013(01):P01010, 2013.

\bibitem{liu2013}
Michael~KP Liu, Natalie Hawkins, Adam~J Ritchie, Vitaly~V Ganusov, Victoria
  Whale, Simon Brackenridge, Hui Li, Jeffrey~W Pavlicek, Fangping Cai, Melissa
  Rose-Abrahams, et~al.
\newblock Vertical {T} cell immunodominance and epitope entropy determine
  {HIV}-1 escape.
\newblock {\em The Journal of clinical investigation}, 123(1):380, 2013.

\bibitem{lee2008}
Ha~Youn Lee, Alan~S Perelson, Su-Chan Park, and Thomas Leitner.
\newblock Dynamic correlation between intrahost {HIV}-1 quasispecies evolution
  and disease progression.
\newblock {\em PLoS computational biology}, 4(12):e1000240, 2008.

\bibitem{da2012}
Jack da~Silva.
\newblock The dynamics of {HIV}-1 adaptation in early infection.
\newblock {\em Genetics}, 190(3):1087--1099, 2012.

\bibitem{maree2000}
Athanasius~FM Mar{\'e}e, Wilco Keulen, Charles~AB Boucher, and Rob~J De~Boer.
\newblock Estimating relative fitness in viral competition experiments.
\newblock {\em Journal of virology}, 74(23):11067--11072, 2000.

\bibitem{leviyang2013}
Sivan Leviyang.
\newblock Computational inference methods for selective sweeps arising in acute
  {HIV} infection.
\newblock {\em Genetics}, 2013.

\bibitem{feeney2005}
Margaret~E. Feeney, Yanhua Tang, Katja Pfafferott, Kathleen~A. Roosevelt, Rika
  Draenert, Alicja Trocha, Xu~G. Yu, Cori Verrill, Todd Allen, Corey Moore,
  Simon Mallal, Sandra Burchett, Kenneth McIntosh, Stephen~I. Pelton, M.~Anne
  St.~John, Rohan Hazra, Paul Klenerman, Marcus Altfeld, Bruce~D. Walker, and
  Philip J.~R. Goulder.
\newblock {HIV}-1 viral escape in infancy followed by emergence of a
  variant-specific {CTL} response.
\newblock {\em The Journal of Immunology}, 174(12):7524--7530, 2005.

\bibitem{strelkowa2012}
Natalja Strelkowa and Michael L{\"a}ssig.
\newblock Clonal interference in the evolution of influenza.
\newblock {\em Genetics}, 192(2):671--682, 2012.

\bibitem{de1999}
M~de~Visser, Clifford~W Zeyl, Philip~J Gerrish, Jeffrey~L Blanchard, Richard~E
  Lenski, et~al.
\newblock Diminishing returns from mutation supply rate in asexual populations.
\newblock {\em Science}, 283(5400):404--406, 1999.

\bibitem{miralles1999}
Rosario Miralles, Philip~J Gerrish, Andr{\'e}s Moya, and Santiago~F Elena.
\newblock Clonal interference and the evolution of {RNA} viruses.
\newblock {\em Science}, 285(5434):1745--1747, 1999.

\bibitem{lang2013}
Gregory~I Lang, Daniel~P Rice, Mark~J Hickman, Erica Sodergren, George~M
  Weinstock, David Botstein, and Michael~M Desai.
\newblock Pervasive genetic hitchhiking and clonal interference in forty
  evolving yeast populations.
\newblock {\em Nature}, 500:571--574, 2013.

\bibitem{zanini2013}
Fabio Zanini and Richard~A Neher.
\newblock Deleterious synonymous mutations hitchhike to high frequency in
  {HIV}-1 env evolution.
\newblock {\em arXiv preprint arXiv:1303.0805}, 2013.

\bibitem{fischer2010}
Will Fischer, Vitaly~V Ganusov, Elena~E Giorgi, Peter~T Hraber, Brandon~F
  Keele, Thomas Leitner, Cliff~S Han, Cheryl~D Gleasner, Lance Green, Chien-Chi
  Lo, et~al.
\newblock Transmission of single {HIV}-1 genomes and dynamics of early immune
  escape revealed by ultra-deep sequencing.
\newblock {\em PloS one}, 5(8):e12303, 2010.

\bibitem{zagordi2010b}
Osvaldo Zagordi, Lukas Geyrhofer, Volker Roth, and Niko Beerenwinkel.
\newblock Deep sequencing of a genetically heterogeneous sample: local
  haplotype reconstruction and read error correction.
\newblock {\em Journal of computational biology}, 17(3):417--428, 2010.

\bibitem{schirmer2012}
Melanie Schirmer, William~T Sloan, and Christopher Quince.
\newblock Benchmarking of viral haplotype reconstruction programmes: an
  overview of the capacities and limitations of currently available programmes.
\newblock {\em Briefings in bioinformatics}, 2012.

\bibitem{topfer2013}
Armin T{\"o}pfer, Osvaldo Zagordi, Sandhya Prabhakaran, Volker Roth, Eran
  Halperin, and Niko Beerenwinkel.
\newblock Probabilistic inference of viral quasispecies subject to
  recombination.
\newblock {\em Journal of Computational Biology}, 20(2):113--123, 2013.

\bibitem{zagordi2010}
Osvaldo Zagordi, Rolf Klein, Martin D{\"a}umer, and Niko Beerenwinkel.
\newblock Error correction of next-generation sequencing data and reliable
  estimation of {HIV} quasispecies.
\newblock {\em Nucleic Acids Research}, 38(21):7400--7409, 2010.

\bibitem{prosperi2012}
Mattia~CF Prosperi and Marco Salemi.
\newblock Qure: software for viral quasispecies reconstruction from
  next-generation sequencing data.
\newblock {\em Bioinformatics}, 28(1):132--133, 2012.

\bibitem{zagordi2011}
Osvaldo Zagordi, Arnab Bhattacharya, Nicholas Eriksson, and Niko Beerenwinkel.
\newblock Shorah: estimating the genetic diversity of a mixed sample from
  next-generation sequencing data.
\newblock {\em BMC bioinformatics}, 12(1):119, 2011.

\bibitem{burger1991}
Harold Burger, Barbara Weiser, Kelli Flaherty, Janet Gulla, Phi-Nga Nguyen, and
  Richard~A Gibbs.
\newblock Evolution of human immunodeficiency virus type 1 nucleotide sequence
  diversity among close contacts.
\newblock {\em Proceedings of the National Academy of Sciences},
  88(24):11236--11240, 1991.

\bibitem{keele2008}
Brandon~F Keele, Elena~E Giorgi, Jesus~F Salazar-Gonzalez, Julie~M Decker,
  Kimmy~T Pham, Maria~G Salazar, Chuanxi Sun, Truman Grayson, Shuyi Wang, Hui
  Li, et~al.
\newblock Identification and characterization of transmitted and early founder
  virus envelopes in primary hiv-1 infection.
\newblock {\em Proceedings of the National Academy of Sciences},
  105(21):7552--7557, 2008.

\bibitem{Huang2012}
Weichun Huang, Leping Li, Jason~R. Myers, and Gabor~T. Marth.
\newblock Art: a next-generation sequencing read simulator.
\newblock {\em Bioinformatics}, 28(4):593--594, 2012.

\bibitem{holmes2002}
Edward~C. Holmes and Andrés Moya.
\newblock Is the quasispecies concept relevant to {RNA} viruses?
\newblock {\em Journal of Virology}, 76(1):460--462, 2002.

\bibitem{grenfell2004}
Bryan~T Grenfell, Oliver~G Pybus, Julia~R Gog, James~LN Wood, Janet~M Daly,
  Jenny~A Mumford, and Edward~C Holmes.
\newblock Unifying the epidemiological and evolutionary dynamics of pathogens.
\newblock {\em Science}, 303(5656):327--332, 2004.

\bibitem{kessinger2013}
Taylor~A Kessinger, Alan~S Perelson, and Richard~A Neher.
\newblock Inferring {HIV} escape rates from multi-locus genotype data.
\newblock {\em Frontiers in Immunology}, 4(252), 2013.

\bibitem{neher2010}
Richard~A Neher and Thomas Leitner.
\newblock Recombination rate and selection strength in {HIV} intra-patient
  evolution.
\newblock {\em PLoS computational biology}, 6(1):e1000660, 2010.

\bibitem{batorsky2011}
Rebecca Batorsky, Mary~F Kearney, Sarah~E Palmer, Frank Maldarelli, Igor~M
  Rouzine, and John~M Coffin.
\newblock Estimate of effective recombination rate and average selection
  coefficient for {HIV} in chronic infection.
\newblock {\em Proceedings of the National Academy of Sciences},
  108(14):5661--5666, 2011.

\bibitem{thompson2002}
Julie~D Thompson, Toby Gibson, Des~G Higgins, et~al.
\newblock Multiple sequence alignment using {ClustalW} and {ClustalX}.
\newblock {\em Current protocols in bioinformatics}, 2.3:1--22, 2002.

\bibitem{guindon2003}
St{\'e}phane Guindon and Olivier Gascuel.
\newblock A simple, fast, and accurate algorithm to estimate large phylogenies
  by maximum likelihood.
\newblock {\em Systematic biology}, 52(5):696--704, 2003.

\bibitem{librado2009}
P~Librado and J~Rozas.
\newblock {DnaSP} v5: a software for comprehensive analysis of dna polymorphism
  data.
\newblock {\em Bioinformatics}, 25(11):1451--1452, 2009.

\bibitem{smoot2011}
Michael~E Smoot, Keiichiro Ono, Johannes Ruscheinski, Peng-Liang Wang, and Trey
  Ideker.
\newblock Cytoscape 2.8: new features for data integration and network
  visualization.
\newblock {\em Bioinformatics}, 27(3):431--432, 2011.

\end{thebibliography}
\bibliographystyle{unsrt}

\section*{Tables}
\begin{table}[!ht]
\caption{\textbf{Haplotype reconstruction performance on simulated datasets.} Simulated datasets were named according to the frequency distribution of their population and their nucleotide diversity. Populations with uniform and log-normal frequency distributions are labeled as U and L, respectively, followed by a number denoting the percentage nucleotide diversity between genomes. In all cases there were 9 ``true'' genomes.}
\

      \begin{tabular}{c c c c c c}
        \hline
	{\centering{\textbf{Dataset}}} & \multicolumn{1}{p{1.55cm}}{\centering{\textbf{Nucleotide diversity}}} & \multicolumn{1}{p{1.8cm}}{\centering{\textbf{No. of haplotypes reconstructed}}} & \multicolumn{1}{p{2cm}}{\centering{\textbf{Mean Hamming distance (MHD)}}} & \multicolumn{1}{p{1.5cm}}{\centering{\textbf{MHD of top 4 haplotypes}}} & \multicolumn{1}{p{1.5cm}}{\centering{\textbf{Range Hamming distance}}} \\ \hline
        U2 & 2 & 9 & 4.33 & 2.25 & (1-9) \\
	U4 & 4 & 9 & 4.33 & 2 & (2-9) \\
	U10 & 10 & 9 & 4.66 & 2.25 & (1-9) \\
	L2 & 2 & 5 & 46.2 & 23 & (7-139) \\
	L4 & 4 & 9 & 158.44 & 11 & (1-427) \\
	L10 & 10 & 7 & 283.85 & 65.25 & (46-706) \\ \hline
      \end{tabular}
\end{table}

\begin{table}[h!]
\caption{\textbf{Haplotype reconstruction of quasispecies datasets.} Haplotype reconstructions from populations having a simulated viral quasispecies structure. Nine quasispecies were present in each simulated population and the diversity between them was 4\% (see Methods). The pruned haplotype genomes were created by excluding the initial and final 50 bp from the reconstructed haplotypes to avoid errors due to low read coverage.}
\

      \begin{tabular}{c c c c c c}
        \hline
        {\centering{\textbf{Dataset}}} & \multicolumn{1}{p{1.8cm}}{\centering{\textbf{No. of haplotypes reconstructed}}} & \multicolumn{1}{p{2cm}}{\centering{\textbf{Mean Hamming distance (MHD)}}} & \multicolumn{1}{p{1.5cm}}{\centering{\textbf{Range Hamming distance}}} & \multicolumn{1}{p{1.5cm}}{\centering{\textbf{MHD of top 4 haplotypes}}} & \multicolumn{1}{p{1.5cm}}{\centering{\textbf{MHD of top 4: pruned}}} \\ \hline
	LQ4\_1 & 7 & 79.14 & (8-225) & 15.75 & 3.75 \\
	LQ4\_2 & 8 & 137.5 & (15-403) & 20.5 & 6.5 \\
	LQ4\_3 & 7 & 271.57 & (15-1086) & 60 & 49 \\
	LQ4\_4 & 7 & 86.71 & (10-261) & 14.75 & 3.75 \\
	LQ4\_5 & 7 & 62.14 & (8-257) & 12.25 & 1.75 \\
	LQ4\_6 & 6 & 51.5 & (13-231) & 13.5 & 4 \\
	LQ4\_7 & 7 & 101.71 & (14-344) & 20 & 6 \\
	LQ4\_8 & 6 & 104 & (8-305) & 41.25 & 13.5 \\
	LQ4\_9 & 6 & 58.5 & (13-194) & 17 & 5.75 \\
	LQ4\_10 & 8 & 134.63 & (9-349) & 16 & 2.25 \\ \hline
      \end{tabular}
\end{table}
\samepage
\clearpage
\section*{Figures}
\begin{figure}[h!t]
  \begin{center}
  \includegraphics[width=5in]{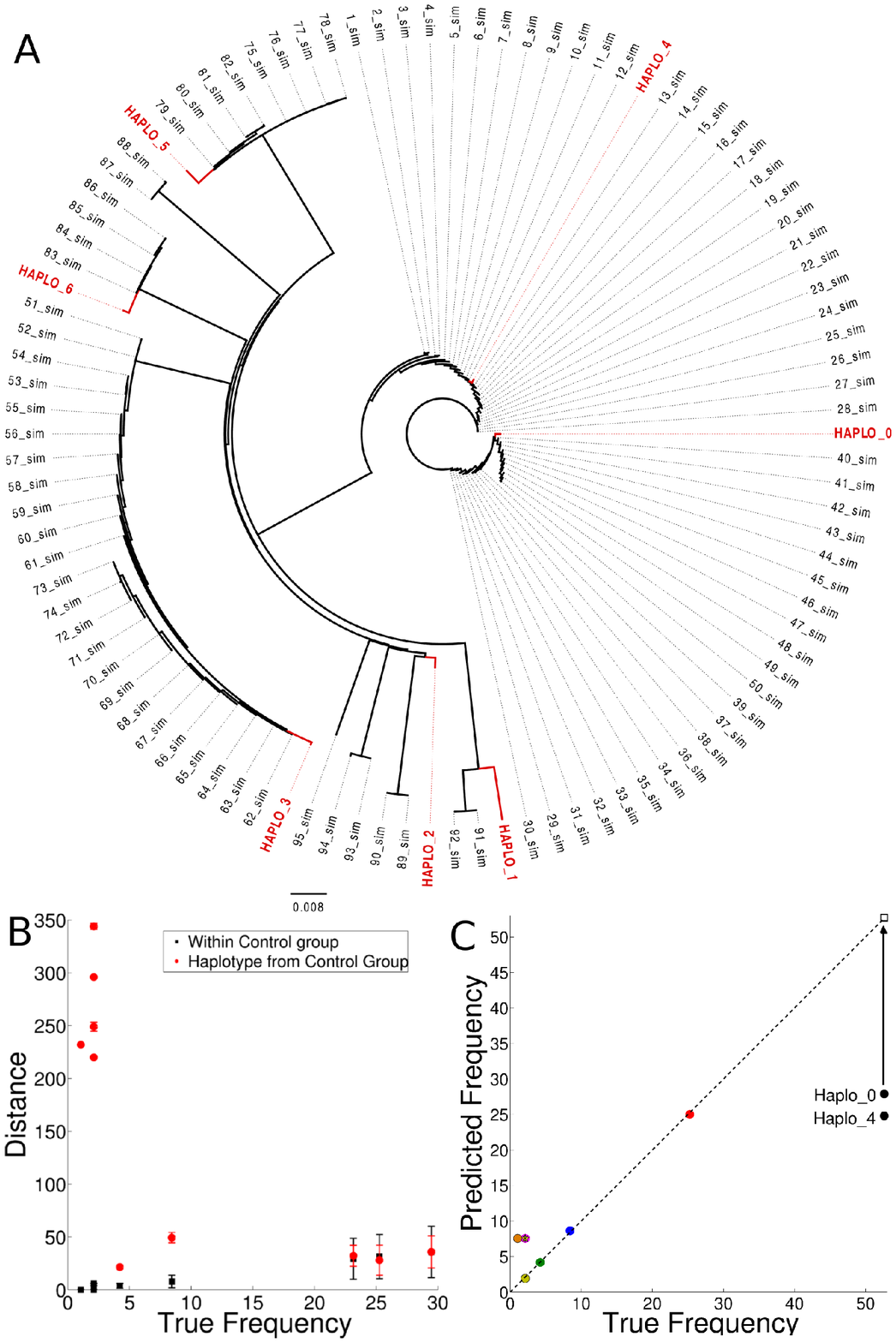}
  \end{center}
\end{figure}

\begin{figure}
  \caption{We simulated HIV-1 population (LQ4\_1; Table 2) with 9 quasispecies consisting of 50, 24, 8, 4, 2, 2, 2, 2 and 1, sequences respectively. The genome sequences within each quasispecies contained at most 0.1$\%$ nucleotide diversity (see Methods). (A) Seven reconstructed whole genome haplotypes (red) were aligned with the 95 simulated genomes. Dark lines represent the branches of the phylogentic tree (the dotted lines pointing outwards provide better visualization). (B) Average Hamming distance within each quasispecies (black), and between the reconstructed haplotype and the quasispecies at minimum Hamming distance (red). Because one haplotype can represent multiple quasispecies, we used the phylogenetic tree to determine the genomes represented by a haplotype. Markers represent the mean Hamming distance and error bars represent 1 standard deviation. (C) The haplotype frequencies (vertical axis) predicted from haplotype reconstruction pipeline are plotted against the true quasispecies frequencies (horizontal axis). The dominant quasispecies was reconstructed as two haplotypes (Haplo\_0 and Haplo\_4) and summation of their predicted frequencies is indicated by an open square. The haplotype Haplo\_2 (predicted frequency of 7.55\%) represents four low prevalence quasispecies: one quasispecies with true frequency of 1.05\% (orange circle) and three with true frequencies of 2.1\% (see the overlapping points next to orange circle).The points lying on the dashed diagnal line represent perfect haplotype frequency predictions.}
\end{figure}

\begin{figure}[!ht]
  \begin{center}
  \includegraphics[width=4.5in]{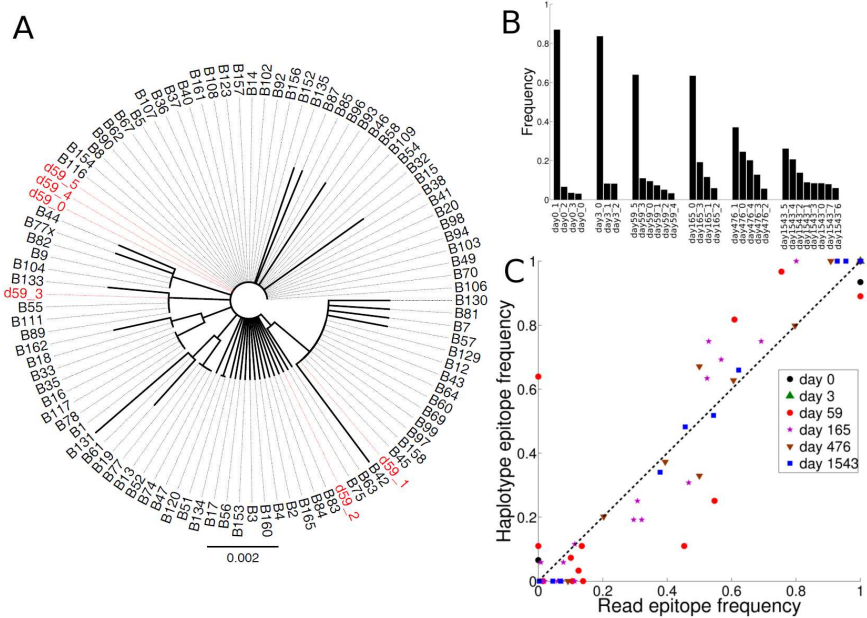}
  \end{center}
  \caption{Validation of reconstructed haplotypes from subject 9213. (A) Neighbor-joining phylogenetic tree generated from the global alignment of the \textit{vif} genes extracted from the 6 haplotypes reconstructed from the day 59 data (red) and the SGA \textit{vif} gene sequences (black) obtained from \cite{henn2012}. (B) The predicted haplotype frequencies for subject 9213 plotted for each sample (day 0, day 3, day 59, day 165, day 476, and day 1543). Note that the variation in read coverage (667.7, 724.4, 750.5, 299.7, 227.6 and 540.7, respectively) does not seem to influence the number of predicted haplotypes. (C) Comparison of the epitope frequencies estimated from reconstructed haplotypes and those from the local analysis of sequencing reads described in \cite{henn2012}. Markers represent different days. Variants of each epitope are represented as different observations. Several epitope variants with 100\% read frequency and a 100\% predicted frequency in reconstructed haplotypes appear as overlapping points in the upper right corner on the diagonal. The overlapping points were not considered for the correlation analysis. The Spearman's rank correlation coefficient between predicted epitope frequencies extracted from complete genome haplotypes and the local read frequencies was $\rho=0.85$ (p-value $< 10^{-15}$).}
\end{figure}

\begin{figure}[!ht]
  \begin{center}
  \includegraphics[width=5in]{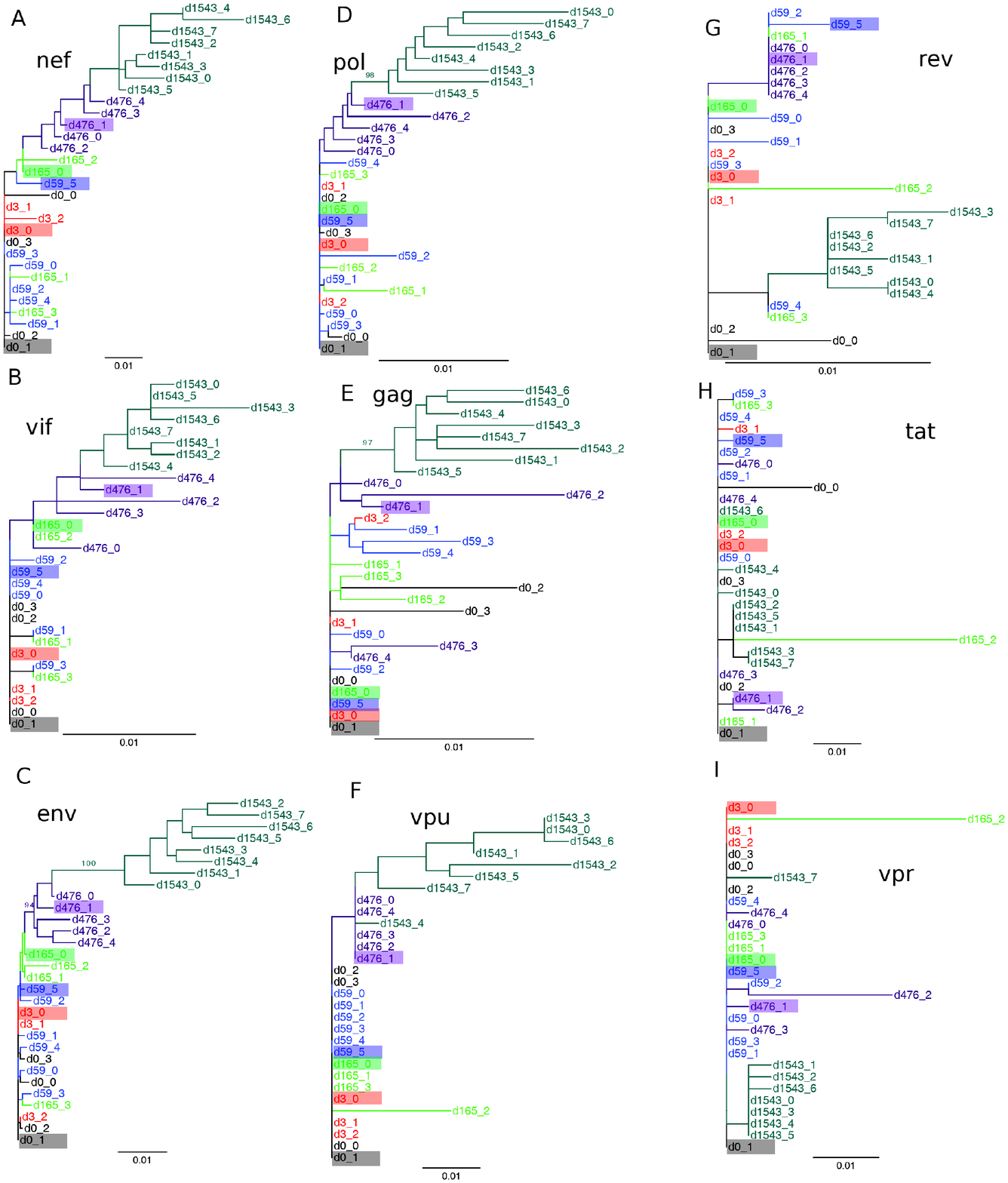}
  \end{center}
  \caption{Phylogenetic trees of the reconstructed HIV-1 genes extracted from genome haplotypes. The phylogenetic trees generated from the nucleotide sequences of the (A) \textit{nef}, (B) \textit{vif}, (C) \textit{env}, (D) \textit{pol}, and (E) \textit{gag} genes, that are targeted by one or more CTLs \cite{henn2012} exhibit a strong to intermediate signal for temporal selection. The \textit{vpu} (F), \textit{rev} (G), \textit{tat} (H), and \textit{vpr} (I) genes exhibit weak or no signal for temporal selection in their phylogenetic trees. The maximum likelihood (ML) phylogenetic trees were reconstructed with PHYML using the GTR nucleotide substitution model, gamma distributed rate variation across sites and 6 substitution rate categories (see Methods for details). 100 bootstrap replicates were performed and support values of $>=$90 are shown on the corresponding branch. Colors denote different temporal haplotypes: day 0 (grey), day 3 (red), day 59 (blue), day 165 (green), day 476 (purple) and day 1543 (teal). The dominant haplotype for each sample is highlighted by a colored box.}
\end{figure}

\begin{figure}[!ht]
  \begin{center}
  \includegraphics[width=3in]{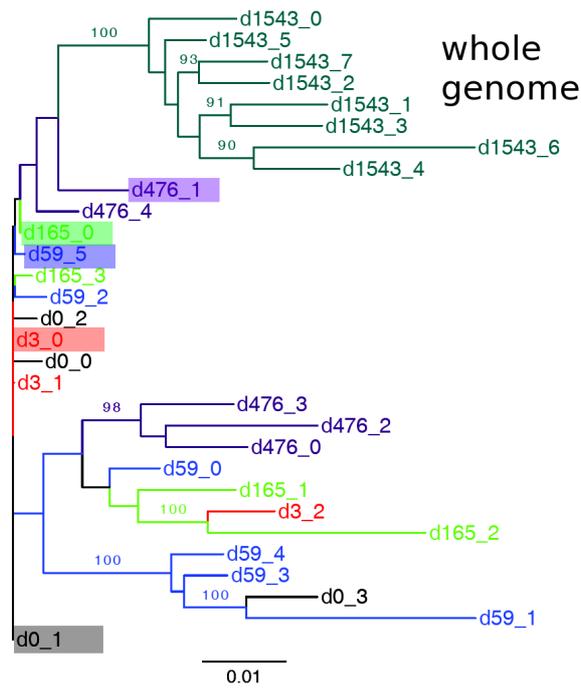}
  \end{center}
  \caption{Phylogenetic tree for complete genome haplotypes. A whole genome nucleotide phylogenetic tree was generated using the reconstructed haplotypes. The ML phylogenetic tree was made with PHYML using the GTR nucleotide substitution model, gamma distributed rate variation across sites and 9 substitution rate categories (see Methods for details). 100 bootstrap replicates were performed and support values of $>=$90 are shown on the corresponding branch. Colors as in Figure 3.}
\end{figure}

\begin{figure}[!ht]
  \begin{center}
  \includegraphics[width=4in]{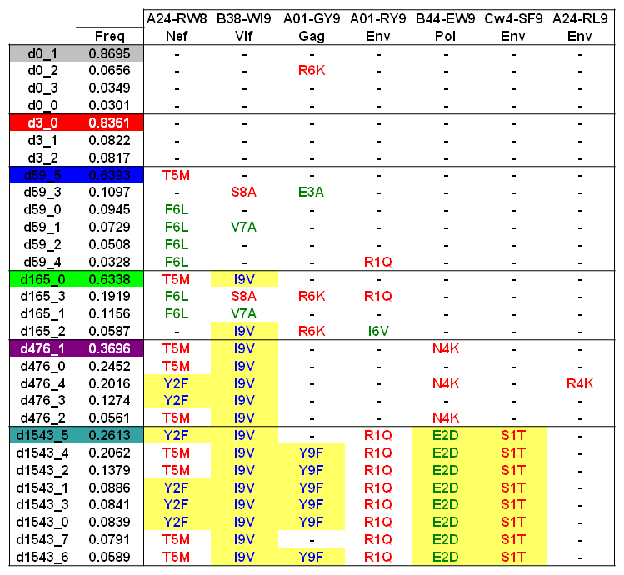}
  \end{center}
  \caption{Epitope escapes in complete genome haplotypes. The escape mutations acquired in the seven epitopes identified by Henn et al. \cite{henn2012} are given per reconstructed haplotype for each time point. Haplotypes reconstructed from each temporal sample are given in a descending order of their frequencies. The dominant haplotype for each time point is depicted by the same colors as in Figure 3. Multiple escapes for the same epitope are depicted by different colors. Escape mutations for each epitope are denoted by a 3-letter code with the first letter as the original epitope amino acid, second letter as the position in the epitope and the third letter as the mutated amino acid. Mutations at anchor residues are highlighted.}
\end{figure}

\begin{figure}[!ht]
  \begin{center}
  \includegraphics[width=4.5in]{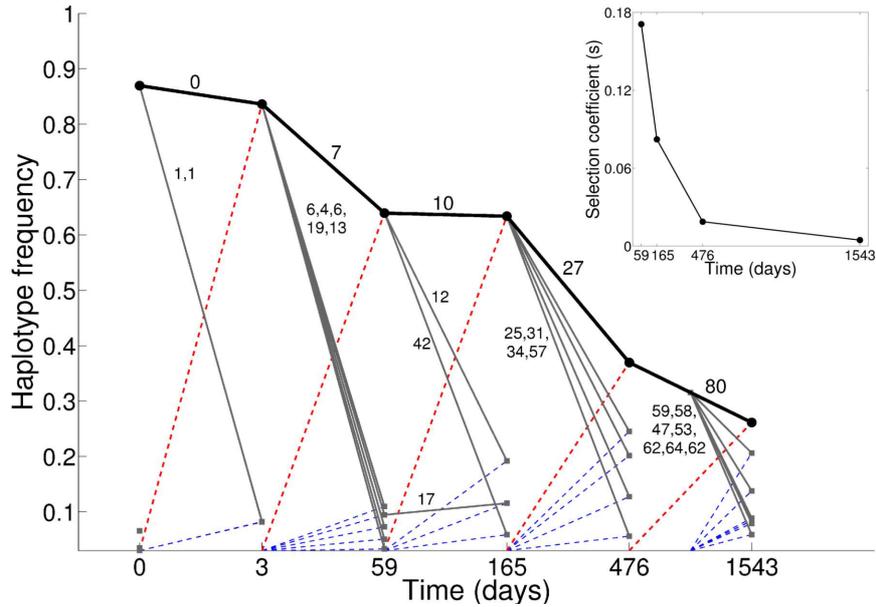}
  \end{center}
  \caption{Selection dynamics of complete genome haplotypes. Dominant haplotypes for each sample are denoted by black circles, and the less abundant haplotypes are denoted by grey squares. The dominant and less abundant haplotypes are connected to those haplotypes from the previous time point having the minimum non-synonymous Hamming distance (mentioned above each edge). The less prevalent haplotypes at day 1543 were closer to the dominant haplotype at the same day compared to any haplotype present at day 476, and thus are shown to have a predecessor at an intermediate time. The rate of replacement by a haplotype genome is represented by dashed lines: in red for dominant haplotypes, and in blue for non-dominant haplotypes. Note that the horizontal axis has no true time scale. Selection coefficients for dominant haplotypes are shown in the inset. Selection coefficients for the dominant haplotype genomes were calculated using the predicted haplotype frequency at time $t$ and its frequency at preceding time point $t-1$ using the method described in \cite{maree2000}. The mutant was considered to have a prevalence of 3\% at the preceeding time point (because we cannot reliably detect haplotypes below a 3\% frequency).}
\end{figure}

\begin{figure}[!ht]
  \begin{center}
  \includegraphics[width=4.8in]{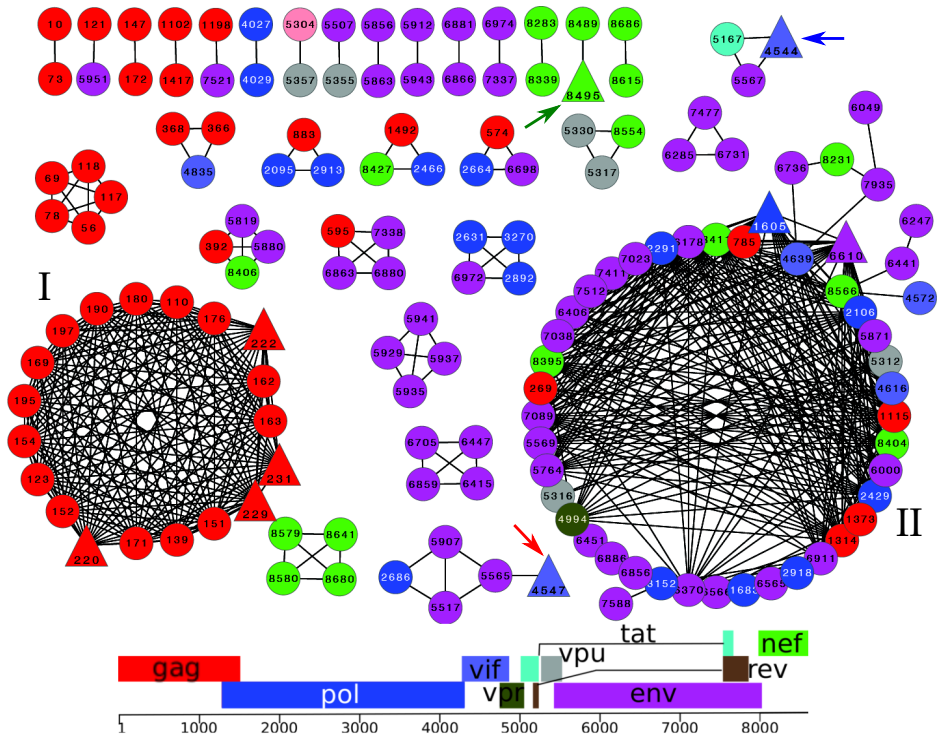}
  \end{center}
  \caption{Network of linked genomic sites. Genomic sites with significant linkage disequilibrium are plotted in the network as nodes. An edge between two nodes represent that the two sites are linked. Sites belonging to different genes are shown in different colors: \textit{gag} as red, \textit{pol} as blue with white fonts, \textit{vif} as violet, \textit{vpr} as dark green, \textit{tat} as light blue, \textit{vpu} as grey, \textit{env} as purple and \textit{nef} as light green. No linked sites were found for \textit{rev}. The legend shows the genomic position and colors assigned to the nine HIV-1 genes. Sites belonging to CTL epitope are shown as triangles. Green arrow indicates the Nef A24-RW8 epitope escape, and blue and red arrows indicate Vif B38-WI9 epitope escapes. ``I'' corresponds to the large cluster containing linked sites from the \textit{gag} gene only, and ``II'' corresponds to the large cluster containing linked sites from several HIV-1 genes.}
\end{figure}

\clearpage
\section*{Supplementary}
\setcounter{figure}{0}
\renewcommand{\figurename}{Supplementary Figure}
  \begin{figure*}[!ht]
  \begin{center}
  \includegraphics[width=4.8in]{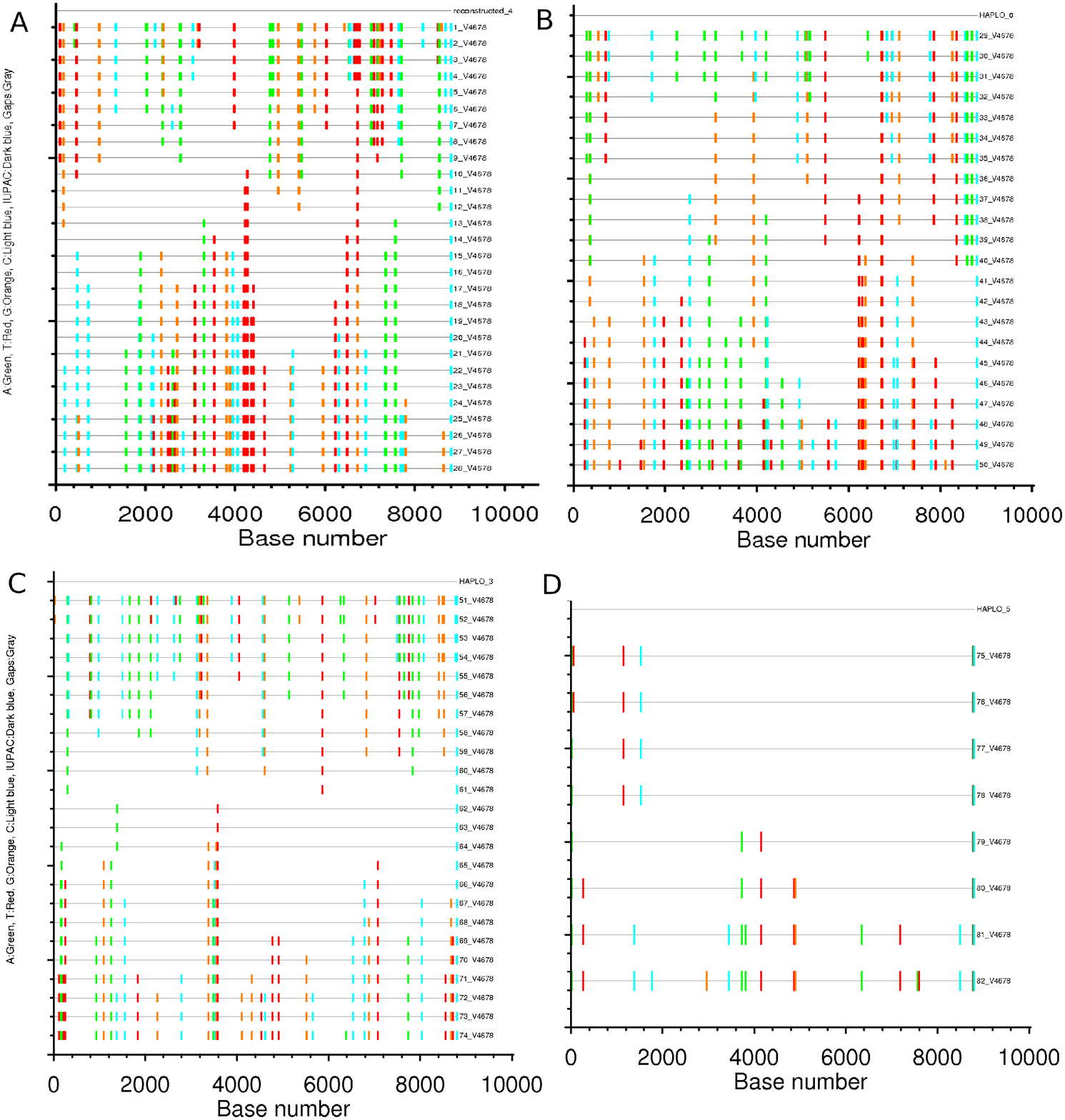}
  \end{center}
  \caption{Sequence alignments of the reconstructed haplotype and the corresponding quasispecies. Sequence alignment of four most prevalent haplotypes (A) haplotype\_4, (B) haplotype\_0, (C) haplotype\_3, and (D) haplotype\_5 and the corresponding quasispecies. Mismatches between the simulated genomes and the reconstructed haplotype are indicated as vertical lines with different colors corresponding to different nucleotides (A$:$ Green, T$:$ Red, G$:$ Orange, C$:$ Light blue and Gaps$:$ Gray). The variation at each genomic site was present only in a subset of genomes demonstrating that the haplotypes are reconstructed as consensus sequences of the corresponding quasispecies.}
\end{figure*}

\end{document}